\documentclass[chicago,numberedappendix,tighten]{emulateapj}

\usepackage{amsmath,amssymb,natbib,graphicx}
\usepackage{color}
\usepackage{ulem}

\slugcomment{DRAFT \today}

\shorttitle{Baryon Loaded Relativistic Blastwaves}
\shortauthors{Chakraborti \& Ray}

\begin{document}


\title{Baryon Loaded Relativistic Blastwaves in Supernovae}


\author{Sayan Chakraborti\altaffilmark{1} \& Alak Ray\altaffilmark{1}}
\affil{Department of Astronomy and Astrophysics, Tata Institute of Fundamental Research,\\
    1 Homi Bhabha Road, Mumbai 400 005, India}

\email{sayan@tifr.res.in, akr@tifr.res.in}


\altaffiltext{1}{Also at Institute for Theory and Computation,
Harvard-Smithsonian Center for Astrophysics, 60 Garden St.,
Cambridge, MA 02138, USA}


\begin{abstract}
We provide a new analytic blastwave solution which generalizes the
Blandford-McKee solution to arbitrary ejecta masses and Lorentz factors.
Until recently relativistic supernovae have been discovered only through their
association with long duration Gamma Ray Bursts (GRB). The blastwaves of such
explosions are well described by the Blandford-McKee (in the ultra relativistic regime)
and Sedov-Taylor (in the non-relativistic regime) solutions during their
afterglows, as the ejecta mass is negligible in comparison to the swept up mass.
The recent discovery of the relativistic supernova SN 2009bb, without
a detected GRB, opens up the possibility of highly baryon loaded mildly relativistic
outflows which remains in nearly free expansion phase during the radio
afterglow. In this work, we consider a massive, relativistic shell,
launched by a Central Engine Driven EXplosion (CEDEX), decelerating adiabatically
due to its collision with the pre-explosion circumstellar wind profile of the
progenitor. We compute the synchrotron emission from relativistic
electrons in the shock amplified magnetic field. This models
the radio emission from the circumstellar interaction of a CEDEX. We show that
this model explains the observed radio evolution of the prototypical SN 2009bb and
demonstrate that SN 2009bb had a highly baryon loaded, mildly relativistic outflow.
We discuss the effect of baryon loading on the dynamics and observational
manifestations of a CEDEX. In particular, our predicted angular size of SN 2009bb
is consistent with VLBI upper limits on day 85, but is presently
resolvable on VLBI angular scales, since the
relativistic ejecta is still in the nearly free expansion phase. 
\end{abstract}





\section{Introduction}
Ultra relativistic bulk motion of matter particles in astrophysical
settings is implied most notably in Gamma Ray Bursts (GRBs) (see
\citet{1999PhR...314..575P,2004RvMP...76.1143P} for reviews). Afterglows from GRBs
are generated from 
the emission by relativistic shocks that result from slowing down of a relativistic
shell by the the medium surrounding the progenitor star that exploded.
Similar interaction of stellar material (ejecta) from an exploding star
with the circumstellar matter (CSM) gives rise to non-relativistic shocks in
core collapse supernovae. 

Fluid dynamical treatment of ultra-relativistic spherical blast waves mediated by strong 
shocks\footnote{\label{strong}Here the strong shocks are those where
the random kinetic energy per particle behind the shock is much greater than the unshocked
medium, i.e., $p_2/n_2 \gg p_1/n_1$, the subscripts 2 and 1 denote the shocked gas and
unshocked gas respectively and $p, n$ are the pressure and number densities.}
has been given by \citet{1976PhFl...19.1130B,1977MNRAS.180..343B}. They describe a
similarity solution of
an explosion of a fixed amount of energy in a uniform medium. This includes an adiabatic
blast wave and an impulsive injection of energy on a short timescale as well as an
explosion where the total energy increases with time, suggesting that the blast wave
has a continuous central power supply. Another important model
considered by them is that of a blast wave propagating into a spherically symmetric wind.
On the other hand, the initial nearly free expansion of a non-relativistic supernova
blast wave, interacting with the surrounding circumstellar medium, was found by
\citet{1982ApJ...259..302C,1985Ap&SS.112..225N}. Once the
blast wave sweeps up more CSM material than its own rest mass, the self-similar
solutions of non-relativistic blast waves are described in the
Newtonian regime by the \citet{Sedov} \citet{vonNeumann} \citet{1950RSPSA.201..159T}
solution.

In this paper we provide an analytic solution of the standard model of relativistic
hydrodynamics \citep[see e.g.][]{1999PhR...314..575P,1999ApJ...512..699C}
for an adiabatic blastwave.
Here, the exploding shell
decelerates due to inelastic collision with an external medium. That 
is, we provide the solution for an arbitrary Lorentz factor of the
expanding supernova shell. 
The need for such a solution which can handle a trans-relativistic outflow is motivated 
by the discovery of SN 2009bb, a type Ibc supernova without a detected GRB which
shows clear evidence of a mildly relativistic outflow powered by a central engine 
\citep{2010Natur.463..513S}. SN 2009bb-like
objects (Central Engine Driven Explosions, hereafter CEDEX)
differ in another significant way from classical GRBs: they are highly baryon
loaded explosions with non-negligible ejecta masses. 
Our new analytic
blastwave solution therefore generalizes the \citet{1976PhFl...19.1130B} result, 
in particular their impulsive, adiabatic blast wave in a
wind-like $\rho\propto r^{-2}$ CSM.

The new class of relativistic supernovae without a detected GRB, e.g. SN 2009bb,
relaxes two important and well-known constraints of the GRBs, namely the 
\textit{compactness problem} and \textit{baryon contamination}. Thus highly baryon
loaded mildly relativistic outflows can remain in the nearly free expansion
phase during the radio afterglow. Therefore we consider the
evolution of a massive relativistic shell launched by a CEDEX,
experiencing collisional slowdown due to interaction with the pre-explosion
circumstellar wind of the progenitor.
We calculate
the time evolution of the radius, bulk Lorentz factor and thermal energy of the
decelerating blastwave. Our solution reduces to the \citet{1976PhFl...19.1130B}
solution, in the ultra-relativistic, negligible
ejecta mass limit. We also quantify the density of accelerated electrons,
amplified magnetic fields and hence the radio synchrotron emission from
the blastwave of a CEDEX.


In Section \ref{golpo} we outline the conditions that require
ultra-high bulk Lorentz factors and very low ejecta masses in classical Long GRBs. We
argue why these constraints are relaxed in the case of mildly relativistic outflows
detected in SN 2009bb like events, that have no detected GRBs.
It provides the motivation for the analytic solution derived here.
In Section \ref{blast} we develop the analytic solution of a relativistic blast wave
launched by a CEDEX, for the collisional slowdown model described by
\citet{1999PhR...314..575P}. We use this solution to show that the SN 2009bb
blast wave is substantially baryon loaded and remains in the nearly free expansion
phase throughout the $\sim1$ year of observations (Figure \ref{R_t}). 
In section \ref{energy} we discuss the
relativistic blast wave energetics. We quantify the amount of shock accelerated
electrons and magnetic field amplification in a CEDEX blast wave.
In Section \ref{spectrum} we use this information to model the radio spectrum
and light curve of a CEDEX in the nearly free expansion phase.
In section \ref{inversion} we provide expressions to deduce the blast
wave parameters from the observed radio spectrum.
We also compare our blastwave solution
to other known solutions relevant for relativistic blastwaves
in the literature (Section \ref{comp}).
In section \ref{polao}
we discuss the implications of baryon loading in determining the evolution
and observational signatures of a CEDEX.
In the Appendix we provide the analytic expressions for the temporal evolution
of the blast wave parameters (Section \ref{exact}). We demonstrate that our
solution reduces to the ultra-relativistic \cite{1976PhFl...19.1130B} solution
for a constant velocity wind, in the low mass, ultra-relativistic limit
(Section \ref{ultra}). In Section \ref{radiotv} we provide 
``stir-fry expressions''\footnote{\label{fry}We note that
a similar term ``TV Dinner Equations" appeared in
the Astrophysical literature in a paper on GRBs \citep{1999ApJ...525..737R}
to denote equations where ``numerical values for physical constants have been inserted, 
so they are ready to use without further preparation"; in the present work, ``stir-fry"
expressions would allow users to pick and choose usable formulae
derived or approximated in this work and quickly (re)assemble their own combination
of useful input parameters.} 
for the radio spectrum of a CEDEX with known initial blast wave
parameters. In Section \ref{inversiontv} we invert the problem and provide
handy expressions for estimating the initial parameters for a CEDEX from
radio observations. In Section \ref{vlbi} we provide an expression to
compute angular size of a CEDEX from multi-band radio spectrum. Using the radio
spectrum of SN 2009bb, we show that our predicted angular size is consistent
with the reported upper limits \citep{2010arXiv1006.2111B} from VLBI, but should
be resolvable presently.

\section{Relativistic outflow in GRBs and CEDEX}
\label{golpo}
Relativistic supernovae have, until recently, been discovered only through
their temporal and spatial association with long duration GRBs.
An acceptable model for GRBs must find a way to circumvent the
\textit{compactness} and \textit{baryon contamination} problems. The observed
rapid temporal variability in GRBs imply a very compact source. The compactness
problem pointed
out by \citet{1975NYASA.262..164R} and \citet{1978Natur.271..525S}, indicates that
$\gamma$-ray photons of sufficiently high energies will produce electron-positron
pairs and will not be able to come out due to the resulting high opacity. 
However, the observation of high energy
photons from GRBs was reconciled with such theoretical constraints
by \citet{1986ApJ...308L..47G} and
\citet{1986ApJ...308L..43P}, allowing the high energy photons to come out from a
relativistic explosion. This requires initial bulk Lorentz factors of
the radiating shell, $\gamma_0\gtrsim10^2$ \citep{1999PhR...314..575P}. 
\citet{1990ApJ...365L..55S}
pointed out that in the presence of even small amounts of baryons or
\textit{baryon contamination}, essentially the entire energy of the explosion
gets locked up in the kinetic energy of the baryons, leaving little energy for
the electromagnetic display. This problem was solved by \citet{1992MNRAS.258P..41R}
and independently by several authors
\citep{1992ApJ...395L..83N,1994ApJ...430L..93R,1994ApJ...427..708P}, by considering
the reconversion of the kinetic energy of the fireball into radiation, due to
interaction with an external medium or via internal shocks. 
However
the allowed initial mass is still very small ($M\approx10^{-6}M_\odot$), as too much baryonic
mass will slow down the explosion and it will no longer be relativistic
\citep{1999PhR...314..575P}. Hence, observation of a short bright pulse of
$\gamma$-ray photons in GRBs require a very small amount of mass to be ejected with
a very high bulk Lorentz factor.

The burst in the GRB itself
results from the conversion of kinetic energy of ultra-relativistic particles or
possibly the electromagnetic energy of a Poynting flux to radiation in an optically
thin region. An inner engine is believed to accelerate the outflow to relativistic speeds,
although the engine may remain hidden from direct observations.
The ``afterglow" on the other hand results from the slowing down of a relativistic
shell on the external medium surrounding the progenitor star. There can also be an additional
contribution to the afterglow from the inner engine that powers the GRB, since the
engine may continue to emit energy for longer duration with a lower intensity and may
produce the earlier part of the afterglow, say the first day or two in GRB 970228 and
GRB 970508 \citep{1999PhR...314..575P,1998PhRvL..80.1580K}.

In supernovae associated with GRBs, after a brief high energy electromagnetic display,
the relativistic ejecta continues to power a long lived radio afterglow
\citep{1998Natur.395..663K,2004Natur.430..648S}. Since the emergence of $\gamma$-ray
photons, early in its evolution, constrains the initial relativistic ejecta mass to
be very small and the initial bulk Lorentz factor to be very large, the relativistic
ejecta sweeps up more circumstellar material than its own rest mass by the time the
radio afterglow is detected.
The evolution of the radiative blastwave has been described by \citet{1998ApJ...509..717C}.
During the radio afterglow phase if radiative losses do not take away a significant
fraction of the thermal energy, the blastwave may be treated as adiabatic. Under these
conditions, the evolution of the blastwave is well described by the
\citet{1976PhFl...19.1130B} solution if the blastwave remains ultra-relativistic
or by the Sedov-Taylor solution
if the blastwave has slowed down into the Newtonian regime. The interaction
of GRBs with their circumstellar wind has been discussed by \citet{2000ApJ...536..195C}.
The spectra and
light curves of GRB afterglows have been computed in the ultra-relativistic regime
by \citet{1998ApJ...497L..17S} and in the Newtonian regime by \citet{2000ApJ...537..191F}.

\citet{2010Natur.463..513S} have recently discovered bright radio emission, 
associated with the type Ibc SN 2009bb, requiring a substantial relativistic
outflow powered by a central engine. The search for a $\gamma$-ray
counterpart, spatially and temporally coincident with SN 2009bb, in data
obtained from the all-sky Inter Planetary Network (IPN) of high energy
satellites, did not detect a coincident GRB \citep{2010Natur.463..513S}.
The absence of detected $\gamma$-rays from this event relaxes the constraints
of very high Lorentz factors and very low mass in the relativistic ejecta.
In fact, the radio emitting outflow in SN 2009bb is mildly relativistic
\citep{2010Natur.463..513S} and has a large baryonic mass coupled to it, as
evidenced by the nearly free expansion for $\sim 1$ year (Figure \ref{R_t}).
Since the swept up mass is still smaller than the rest mass of the relativistic
ejecta, the evolution of the blastwave can neither be described by the the rapidly
decelerating \citet{1976PhFl...19.1130B} solution nor the Sedov-Taylor solution.
\citet{1998NewA....3..157D,1999ApJ...512..699C} have considered
the emission from a collisionally decelerating blast wave as it transitions from
nearly free expansion to the \citet{1976PhFl...19.1130B} self similar solution.
\citet{1999MNRAS.309..513H} have discussed the subsequent transition to a
Sedov-like phase.

Hence, the discovery of relativistic supernovae, such as SN 2009bb, without
detected GRBs, relaxes the constraints from the \textit{compactness problem}
and \textit{baryon contamination}. This motivates the study of highly baryon
loaded mildly relativistic outflows which can remain in the nearly free expansion
phase during the radio afterglow.

\section{Relativistic Blastwave Solution}
\label{blast}
We use the simple collisional model described by \citet{1999PhR...314..575P,1999ApJ...512..699C}
where the relativistic ejecta forms a shell which decelerates through
infinitesimal inelastic collisions with the circumstellar
wind profile. The initial conditions are characterized by the
rest frame mass $M_0$ of the shell launched by a CEDEX and its initial Lorentz
factor $\gamma_0$. In contrast to the self similar solutions,
which describe the evolution away from the boundaries of the independent
variables \citep{1972AnRFM...4..285B}
and the need for proper normalization to get the correct total energy,
our set of initial conditions directly fixes the total
energy as $E_0 = \gamma_0 M_0 c^2$.

\subsection{Equations of Motion}
The shell slows down by a sequence of infinitesimal inelastic collisions
with the circumstellar  matter. The swept up circumstellar matter is given by
$m(R)$. Conservation of energy and momentum give us
\citep[see][]{1999ApJ...512..699C,1999PhR...314..575P},
\begin{equation}\label{dgamma}
\frac{d\gamma}{\gamma ^{2}-1} = -\frac{dm}{M},
\end{equation}
and
\begin{equation}
dE=c^2(\gamma -1)dm ,
\label{dE}
\end{equation}
respectively, where $dE$ is the kinetic energy converted into thermal energy, 
that is the energy in random motions as opposed to bulk flow, by
the infinitesimal collision. In the adiabatic case this energy is retained in
the shell and we have the analytic relation \citep[from][]{1999PhR...314..575P}
\begin{align}
\frac{m(R)}{M_{0}}
&=-(\gamma _{0}-1)^{1/2}(\gamma _{0}+1)^{1/2}\\  \nonumber
&\times \int_{\gamma_{0}}^{\gamma }(\gamma' -1)^{-3/2}(\gamma' +1)^{-3/2}d\gamma' .
\label{mgamma1}
\end{align}
At the age of 17 days, the radio luminosity of SN 2009bb was
$L_\nu\sim5\times10^{28}$ ergs s$^{-1}$ Hz$^{-1}$ at 8.4 GHz \citep{2010Natur.463..513S}.
This implies a radio luminosity of $\nu L_\nu\sim4.2\times10^{38}$ ergs s$^{-1}$. Compared
with the energy in relativistic electrons of $1.3\times10^{49}$ ergs
\citep{2010Natur.463..513S}, this gives a radiation timescale of $\sim10^3$ years
at age 17 days, which is very large compared to the age of the object.
Hence, it is safe to compute the dynamics of the object in the adiabatic limit.
We also note that since the available thermal energy initially grows with time
as $E\propto R\propto t$ (shown later) and the luminosity falls of as
$\nu L_\nu\propto t^{-1}$ (shown later) the adiabaticity condition only gets better
with time. However, the radiation timescale would have been same as the
age at $t\sim70$ seconds, which is comparable to the duration of very
Long Soft GRBs and XRFs.

For a circumstellar medium set up by a steady wind, where we expect a profile with
$\rho\propto r^{-2}$, we have $m(R)=AR$ where $A$ is the mass swept up by a sphere
per unit radial distance. $A\equiv\dot{M}/v_{wind}$ can be set up by a steady
mass loss rate of $\dot{M}$ with a velocity of $v_{wind}$ from the pre-explosion
CEDEX progenitor, possibly a Wolf Rayet star.
Note that our definition of $A$ is equivalent to $4\pi q$
in Equation (3 and 5) of \citet{1982ApJ...259..302C}.
Integrating the right hand side then gives us
\begin{equation}
\frac{AR}{M_{0}}=\frac{\gamma\sqrt{\gamma_0^2-1}}{\sqrt{\gamma^2-1}} -\gamma_0 .
\end{equation}
For the physically relevant domain of $\gamma>1$ this equation has only one analytical
root at
\begin{equation}
\gamma=\frac{\gamma_0 M_0 + A R}{\sqrt{M_0^2+2 A \gamma_0 R M_0 + A^2 R^2}} ,
\label{gammaR}
\end{equation}
which gives the evolution of $\gamma$ as a function of $R$.
This is similar to Equation 8 of \citet{1999ApJ...512..699C} for
a adiabatic spherical blastwave in a wind like ($\rho\propto r^{-2}$) CSM profile.
The amount of kinetic
energy converted into thermal energy when the shell reaches a particular $R$
can be obtained by integrating Equation (\ref{dE}) after substituting for $\gamma$
from Equation (\ref{gammaR}) and $dm=AdR$, to get
\begin{align}\label{ER}
E&=c^2 \left(-M_0-A R\right. \\ \nonumber
&\left.+\sqrt{M_0^2+2 A \gamma_0 R M_0+A^2 R^2}\right).
\end{align}

\subsection{Evolution in Observer's Time}
The evolution of $R$ and $\gamma$ can be compared with observations once we have the
time in the observer's frame that corresponds to the computed $R$ and $\gamma$.
For emission along the line of sight from a blastwave with a constant $\gamma$
the commonly used expression \citep{1997ApJ...476..232M} is
\begin{equation}\label{rees}
t_{obs} = \frac{R}{2 \gamma^2 c}.
\end{equation}
However, \citet{1997ApJ...489L..37S} has pointed out that for a decelerating
ultra-relativistic
blastwave the correct $t_{obs}$ is given by the \textit{differential} equation
\begin{equation}
d t_{obs} = \frac{dR}{2 \gamma^2 c} .
\label{dtobs}
\end{equation}
We substitute $\gamma$ from Equation (\ref{gammaR}) and integrate both sides to get
the exact expression
\begin{equation}
t_{obs}=\frac{R (M_0+A \gamma_0 R)}{2 c \gamma_0 (\gamma_0 M_0+A R)} .
\end{equation}
Note that, this reduces to the \cite{1997ApJ...476..232M} expression only in the
case of nearly free expansion and deviates as the shell decelerates.
In the rest of the work we use $t$ to indicate the time $t_{obs}$ in the observer's
frame. Inverting this equation and choosing the physically relevant \textit{growing branch},
gives us the analytical time evolution of the line of sight blastwave radius, as
\begin{align}\label{R}
R&=\frac{1}{2 A \gamma_0} \times \left(-M_0+2 A c \gamma_0 t\right. \\ \nonumber
&\left.+\sqrt{8 A c M_0 t \gamma_0^3+(M_0-2 A c \gamma_0 t)^2}\right) ,
\end{align}
in the ultra-relativistic regime.
This can now be substituted into Equations (\ref{gammaR} and \ref{ER}) to get the
time evolution of the Lorentz factor $\gamma$ and the thermal energy $E$ (see Appendix).
This gives us a complete solution for the blastwave time evolution,
parametrized by the values for $\gamma_0$, $M_0$ and $A$.

\subsection{Series Expansions}
Even though an analytical solution is at hand, it is instructive to look at the
Taylor expansions in time for the relevant blastwave parameters of radius
\begin{equation}\label{RS}
R=2 c \gamma_0^2 t
-\frac{4 \left(A c^2 \gamma_0^3 \left(\gamma_0^2-1\right)\right) t^2}{M_0}+O\left(t^3\right) ,
\end{equation}
Lorentz factor
\begin{equation}
\gamma=\gamma_0
-\frac{2 \left(A c \gamma_0^2 \left(\gamma_0^2-1\right)\right) t}{M_0}+O\left(t^2\right) ,
\label{GtS}
\end{equation}
and thermal energy
\begin{align}
E&=2 A c^3 (\gamma_0-1) \gamma_0^2 t \\ \nonumber
&-\frac{2 A^2 c^4 (3 \gamma_0 - 2) \gamma_0^3 \left(\gamma_0^2-1\right) t^2}{M_0}+O\left(t^3\right) .
\end{align}
Equation (\ref{GtS}) immediately tells us that the blastwave of a CEDEX
starts out in a nearly free expansion phase, but slows down significantly
by the time $t_{dec}$, when the first negative term in the Taylor
expansion for $\gamma$ becomes equal to $\gamma_0$, where
\begin{equation}\label{dec}
t_{dec}=\frac{M_0}{2 \left(A c \gamma_0 \left(\gamma_0^2-1\right)\right)}.
\end{equation}
This signals the end of the nearly free expansion phase and the solution
enters a Blandford McKee like (see Appendix \ref{ultra}) or Sedov like
phase, depending upon the Lorentz factor at that time.

\section{Blastwave Energetics}
\label{energy}
We shall now use the blastwave solution developed in the previous section to
predict the radio evolution of a CEDEX with a relativistic blastwave slowing
down due to circumstellar interaction.
For the prototypical SN 2009bb, the blastwave was only mildly relativistic
at the time of the observed radio afterglow. In the absence of a significant
relativistic beaming, the observer would receive emission from the entire
shell of apparent lateral extent $R_{lat}$ at a time $t_{obs}$ given by
\begin{equation}
dt_{obs} = \frac{dR_{lat}}{\beta \gamma c} ,
\end{equation}
which is valid even in the mildly relativistic regime.
Because of its mildly relativistic outflow, all expressions derived for
SN 2009bb, use the above Equation rather than Equation \ref{dtobs} which
is applicable in the ultra-relativistic case. In either case radio observations
of SN 2009bb measure essentially the transverse $R_{lat}$ not the line of sight $R$.
Integrating term by term gives us the time evolution of the lateral
radius as
\begin{align}\label{RLS}
R_{lat}&=c \sqrt{\gamma_0^2-1} t \\ \nonumber
&-\frac{2 \left(A c^2 \gamma_0^3 \sqrt{\gamma_0^2-1}\right) t^2}{M_0}+O\left(t^3\right) .
\end{align}
The thermal energy available when the shell has moved out to a radius $R$
is given exactly by Equation (\ref{ER}), however it is again convenient to
look at its Taylor expansion,
\begin{align}
E&=A c^2 (\gamma_0-1) R \\ \nonumber
&-\frac{\left(A^2 c^2 \left(\gamma_0^2-1\right)\right) R^2}{2 M_0}+O\left(R^3\right) .
\end{align}

\subsection{Electron Acceleration}
\citet{1998ApJ...497L..17S} give the minimum Lorentz factor $\gamma_m$ of the
shock accelerated electrons as
\begin{equation}
\label{gamma_m}
\gamma_m=\epsilon_e\left(\frac {p-2} {p-1}\right) \frac {m_p} {m_e} \gamma.
\end{equation}
At the mildly relativistic velocities seen in SN 2009bb, the peak synchrotron
frequency of the lowest energy electrons are likely to be below the synchrotron
self absorption frequency. This explains the $\nu^{5/2}$ low frequency behavior
of the spectrum. Hence, considering a electron distribution with an energy spectrum
$N_0 E^{-p}dE$, which we assume for simplicity to be extending from $\gamma_m m_ec^2$
to infinity, filling a fraction $f$ of the spherical volume of radius $R$,
we need an energy
\begin{equation}
E_e=\frac{4 f (\gamma_m m_ec^2)^{1-p} N_0 \pi  R^3}{3 (p-1)}.
\end{equation}
If a fraction $\epsilon_e\equiv E_e/E$ of the available thermal energy goes into
accelerating these electrons, then for the leading order expansion of $E$ in $R$
the normalization of the electron distribution is given by
\begin{equation}\label{N0}
 N_0\simeq\frac{3 A c^2 \epsilon_e (\gamma_0-1) (\gamma_m m_ec^2)^2}{2 f \pi  R^2}, 
\end{equation}
for $p=3$, as inferred from the optically thin radio spectrum of
SN 2009bb \citep{2010Natur.463..513S}.

In the rest of this work we have made the simplifying assumption of a
constant $\epsilon_e$. The physics of shock acceleration is unlikely to change during
time of interest as relevant parameters such as the shock velocity does not change
much. Moreover, we show later in this work that the light curve of a CEDEX is
controlled by the competition between a decreasing synchrotron flux at high
frequencies and a decreasing optical thickness (hence increasing
flux) at low frequencies. The major light curve features of a shifting peak
frequency and a nearly constant peak flux, arise from this effect rather than
micro-physical processes that may change $\epsilon_e$.

\subsection{Magnetic Field Amplification}
If a magnetic field of characteristic strength $B$ fills
the same volume, it needs a magnetic energy of
\begin{equation}
 E_B=\frac{1}{6} B^2 f R^3.
\end{equation}
If a fraction $\epsilon_B\equiv E_B/E$ goes into the magnetic energy density,
then the characteristic magnetic field is given by
\begin{equation}\label{B}
 B\simeq \frac{c}{R} \sqrt{\frac{6 A \epsilon_B (\gamma_0-1)}{f}}
\end{equation}
In the rest of this work, following the argument in the previous sub-section,
we have made the simplifying assumption of a constant $\epsilon_B$.
This explains the observed $B\propto R^{-1}$ behavior (Figure \ref{BR}) seen
in SN 2009bb. These observations strengthens the case for a nearly
constant $\epsilon_B$.
This feature of an explosion within a $\rho\propto r^{-2}$ wind
profile is also seen in several non-relativistic radio supernovae.
The normalization is given by the progenitor mass loss parameter,
initial Lorentz factor, filling factor of the electrons and the efficiency
with which the thermal energy is used in amplifying magnetic fields.
Note that this phase would last only as long as the expansion is nearly
free or $t<t_{dec}$. Therefore the observed $B-R$ relation in SN 2009bb argues
for a $t_{dec}\gtrsim1$ year and hence a CEDEX with a highly baryon loaded outflow.

Note that the highest energy to which a cosmic ray proton can be accelerated is determined
by the $B \; R$ product \citep{1984ARA&A..22..425H,2005PhST..121..147W}.
We have argued elsewhere \citep{2010arXiv1012.0850C} that the mildly relativistic
CEDEX are ideal for accelerating nuclei to the highest energies in sufficient 
volumetric energy injection rates and independent arrival directions to explain the
post GZK cosmic rays.
The above $B \propto R^{-1}$ dependence seen in Fig \ref{BR} implies
that a CEDEX like SN 2009bb will continue to accelerate cosmic rays up to a
nearly constant very high energy for long times, i.e. for the entire time that
this (nearly free expansion) phase lasts. 

\section{Electron Synchrotron Radio Spectrum}
\label{spectrum}
It is expected that the radio emission from a CEDEX will be produced by
synchrotron emission of shock accelerated electrons in the shock amplified
magnetic field quantified in the previous section.
Once we have the evolution of $N_0$ and $B$, we can follow \citet{1998ApJ...499..810C}
to express the Synchrotron Self Absorbed (SSA) radio spectrum from a CEDEX in the optically
thick regime using \citet{1979rpa..book.....R} as
\begin{equation}
 F_\nu=\frac{\pi R^2}{D^2} \frac{c_5}{c_6} B^{-1/2} \left(\frac{\nu}{2c_1}\right)^{5/2},
\end{equation}
where $c_1$, $c_5$ and $c_6$ are constants given by \citet{1970ranp.book.....P} and
$D$ is the distance to the source. Similarly, the optically thin flux is given by
\begin{equation}
 F_\nu=\frac{4 \pi f R^3}{3 D^2} c_5 N_0 B^{(p+1)/2} \left(\frac{\nu}{2c_1}\right)^{-(p-1)/2},
\end{equation}
In mildly relativistic cases such as SN 2009bb, we may receive radiation from the entire
disk projected on the sky, hence $R$ is to be understood as $R_{lat}$. 
Substituting for $N_0$ and $B$ from Equations (\ref{N0} and \ref{B}) and the leading
order expansion for the projected lateral radius (Eq. \ref{RLS}), we have the optically
thick flux as
\begin{align}
F_\nu&\simeq\left(c^2 c_5 (\gamma_0-1) (\gamma_0+1) \pi  \left(\frac{\nu  t}{c_1}\right)^{5/2}\right) \\ \nonumber
&\left/\left(4 c_6 D^2 \left(\frac{24 A\epsilon_B}{\gamma_0 f+f}\right)^{1/4} \right)\right. .
\end{align}
Similarly, the optically thin flux can be expressed as
\begin{align}
F_\nu&\simeq\frac{24 A^2 c^3 c_1 c_5 \epsilon_B \epsilon_e (\gamma_0-1)^2 (\gamma_m m_e c^2)^2}{(\nu t) D^2 f \sqrt{\gamma_0^2-1}}.
\end{align}
These equations together provide the flux density of a CEDEX as a function of time
and frequency.

\subsection{SSA Peak Frequency}
The transition from the optically thick to optically thin regime happens at the
peak frequency $\nu_p$. At a fixed observation frequency this is reached in time
$t_p$. The condition for the SSA peak may be obtained by equating the optically thin
and thick flux, to get
\begin{align}\nonumber
\nu_p t_p &\simeq 2^{23/14} 3^{5/14} \left(c c_6 \epsilon_e \sqrt[4]{\frac{A^9 \epsilon_B^5}{\gamma_0 f+f}} \sqrt{c_1^7 \left(\gamma_0^2-1\right)}\right. \\
&\left.\times(\gamma_m m_e c^2)^2\frac{1}{\pi f (\gamma_0+1)^2}\right)^{2/7}
\label{nuptp}
\end{align}
All quantities on the right hand side are constants or parameters of the problem
which are uniquely fixed for a particular relativistic supernova.
Hence, during the nearly free expansion phase $\nu_p \propto t_p^{-1}$ and the peak
moves to lower and lower frequencies as the plasma expands and becomes optically
thin with time. This behavior is also seen in the radio spectra of SN 2009bb.

The range in the values of $\nu_p t_p$ is then weakly dependent on the initial
bulk Lorentz factor $\gamma_0$ and the mass loss rate, parametrized by $A$.
The above expression can be used to select the frequency and cadence of
radio follow-ups of type Ibc supernovae for detecting CEDEXs.

\subsection{SSA Peak Flux}
The peak flux density can now be obtained by substituting the expression for
$\nu_p t_p$ into that for $F_\nu$, to get,
\begin{align}\nonumber
F_{\nu p}&\simeq\left(2^{19/14} 3^{9/14} c^{19/7} c_5 \epsilon_B^{9/14} \epsilon_e^{5/7} (A f (\gamma_0-1))^{19/14}\right. \\
&\left.\times(\gamma_m m_e c^2)^{10/7} \pi ^{2/7}\right)/\left(c_6^{2/7} D^2 f^2\right) .
\label{Fnup}
\end{align}
This peak flux is obtained by keeping only the leading orders, for almost free expansion,
is nearly a constant for for $t\ll t_{dec}$. As the blastwave slows down, the decay of 
the peak flux can be obtained exactly from the analytical expression for $R(t)$. However
for simplicity a Taylor expansion can be written down as
\begin{equation}
\frac{F_{\nu p}(t)}{F_{\nu p}(0)}=1-\frac{5 \left(A c \gamma_0^3\right) t}{M_0}+O\left(t^{3/2}\right)
\end{equation}
The peak flux may also come down if synchrotron losses for the electrons are
significant. This would require a self consistent modeling of the time
dependent acceleration and synchrotron losses of the relativistic electrons in
the shock amplified magnetic field of a CEDEX blastwave.

The peak flux density $F_{\nu p}$ is then strongly dependent on the
bulk Lorentz factor $\gamma_0$ and the parametrized mass loss rate, $A$.
This suggests a large range in radio fluxes of CEDEXs and hence increased
sensitivity of the Extended Very Large Array (EVLA) should help in detecting
more of these enigmatic objects.

\subsection{Synchrotron Cooling}
\citet{1979rpa..book.....R} give the characteristic synchrotron frequency of an
electron with Lorentz factor $\gamma_e \gg 1$ in a magnetic field $B$ as
\begin{equation}
\label{freq}
\nu(\gamma_e)=\gamma \gamma_e^2 \frac {q_e B} {2 \pi m_e c}.
\end{equation}
\citet{1998ApJ...497L..17S} provide the critical electron Lorentz factor
$\gamma_c$, above which an electron will loose a significant portion of
its energy within the age of the object, as
\begin{equation}
\label{cool}
\gamma_c= \frac {6 \pi m_e c} {\sigma_T \gamma B^2 t} .
\end{equation}
Substituting this into Equation \ref{freq}, we get the synchrotron
cooling frequency as
\begin{equation}
\label{coolfreq}
\nu_c\equiv\nu(\gamma_c)=\frac {18 \pi m_e c q_e} {\sigma_T \gamma B^3 t^2}.
\end{equation}
Substituting the value of SN 2009bb magnetic field at 20 days
from Table 1 of \citet{2010arXiv1012.0850C}, we get the cooling frequency
as $\sim2.6$ THz. Since the magnetic field decays as $B\propto t^{-1}$, the
cooling frequency will grow as $\nu_c\propto t$. Hence, synchrotron
cooling is unimportant for electrons radiating near the SSA peak
frequency $\nu_p\sim7.6$ GHz, and they are indeed in the slow cooling regime
during the timescale of interest.

The scaling relations for $\nu_c$ and $\nu_{p}$ show that they should have
been comparable at around $\sim1$ day and at a frequency of $\sim140$ GHz.
Very early time ($\sim1$ day) millimeter wave observations of CEDEXs are
therefore encouraged. They may reveal the presence of a synchrotron
cooling break in the radio spectrum. This may be used to determine
the magnetic field independent of the equipartition argument
\citep[see][]{2004ApJ...604L..97C}.

\section{Blastwave Model Inversion}
\label{inversion}
The input parameters of the model are specified by $M_0$, $\gamma_0$, $A$,
$\epsilon_e$ and $\epsilon_B$. Under the assumption of equipartition of the
thermal energy between electrons, protons and magnetic fields we have
$\epsilon_e=\epsilon_B=1/3$. We then eliminate $A$ between Equations
(\ref{nuptp} and \ref{Fnup}) to get
\begin{equation}\label{G0}
\gamma_0^2-1=4 \left(\frac{3 c_6^8 \epsilon_B (D^{2} F_{\nu p})^9}{\pi^8 c_5^9 \epsilon_e f (\gamma_m m_e c^2)^2
   }\right)^{2/19}
\left(\frac{c_1}{(\nu_p t_p)c}\right)^{2},
\end{equation}
which gives us the initial Lorentz factor $\gamma_0$ in terms of the early-time
($t\ll t_{dec}$) SSA peak frequency $\nu_p$ and the peak flux $F_{\nu p}$.
In the non-relativistic limit this gives us
\begin{equation}\label{v}
v\simeq2\left(\frac{3 c_6^8 \epsilon_B (D^{2} F_{\nu p})^9}{\pi^8 c_5^9 \epsilon_e f (\gamma_m m_e c^2)^2
   }\right)^{1/19}
\left(\frac{c_1}{\nu_p t_p}\right),
\end{equation}
which is insensitive to the equipartition parameter $\alpha\equiv\epsilon_e/\epsilon_B$
as $v\propto\alpha^{-1/19}$. The inferred $\gamma_0$ has a weak dependence on
$\gamma_m$ and is to be solved self consistently with Equation (\ref{gamma_m})
giving us $\gamma_0\simeq1.16$ for SN 2009bb.
Not doing so will incur an error in estimating the initial bulk Lorentz factor.
Equations (\ref{G0} or \ref{v}) may be used as a
direct test of whether a radio supernova has relativistic ejecta.

The equipartition parameter $\alpha$ for a given CEDEX may be determined directly
from radio observations if a synchrotron cooling break is seen in the broad
band radio spectrum. \citet{2004ApJ...604L..97C} have used this to determine
the magnetic field in SN 1993J and its radial evolution \citep{2004ApJ...612..974C},
independent of the equipartition argument.
Subsequently $A$ (essentially the scaled mass loss rate)
may be obtained by fitting Equation (\ref{B}) to the B-R data (e.g.\ Figure \ref{BR}).
A direct method of estimating $A$ would be to eliminate $\gamma_0$
between equations (\ref{nuptp} and \ref{Fnup}) to get $A\simeq1.2\times10^{12}$
gram cm$^{-1}$. For a typical Wolf Rayet wind velocity of $10^3$ km s$^{-1}$
this density profile may be set up by a constant mass loss rate of
$\dot{M}\simeq1.9\times10^{-6} M_\odot$ yr$^{-1}$. This is consistent with
the mass mass loss rate of the SN 2009bb progenitor already determined
by \citet{2010Natur.463..513S}.

All CEDEXs at a given distance, in their nearly free expansion phase,
are characterized by one of the constant $\gamma_0$ curves (Figure \ref{chev}) and
one of the constant $\dot{M}$ curves (again Figure \ref{chev}). This
$F_{\nu p}$ vs $\nu_p t_p$ diagnostic plot for CEDEXs is essentially the
relativistic analogue of Figure 4 of \citet{1998ApJ...499..810C}
where non-relativistic radio supernovae of different types are seen to lie on
one of the constant velocity curves. For example, SN 2009bb, lies on a intermediate
mass loss parameter $A$ and a mildly relativistic $\gamma_0$. For the same initial
$\gamma_0$ as SN 2009bb, an object encountering lower circumstellar density,
would be significantly fainter and faster evolving, such as those occurring near
the points of intersection of the solid green curve and the dashed red curve.
These objects, possibly at the faint end of the luminosity function of CEDEXs,
may be discovered by a dedicated high sensitivity search for relativistic
outflows from nearby type Ibc supernovae with the EVLA.

The ejecta mass can then be inferred from the deceleration timescale $t_{dec}$
fitting the time evolution of the observed radii with the Taylor expansion
(Equation \ref{RLS}) of $R_{lat}$ obtained from our model.
In Figure \ref{R_t} we compare predictions from our
model with the observed temporal evolution of the blastwave radius. The observed
radii are consistent with a nearly free expansion for an apparent lateral velocity
of $\gamma\beta=0.527\pm0.022 c$. As the ejecta has not yet slowed down significantly,
it is not possible to determine $M_0$ but we can
nevertheless put a lower limit on the ejecta mass.
Models with $M_0$ below $10^{-2.5} M_\odot$ are ruled out at the $98.3\%$ level by
the observations of SN 2009bb. Hence, radio observations of a relativistic supernovae
can constrain all the physically relevant parameters of our model.

\section{Comparison with Known Solutions}
\label{comp}
\citet{1976PhFl...19.1130B} (hereafter BM) 
analyze the dynamics of both non-relativistic (NR)
and ultra-relativistic (ER) blast waves in the adiabatic impulsive (AI) approximation
as well as steady injection (SI) from a central power supply. A classification of the
shock dynamics and the corresponding synchrotron and inverse Compton radiation
from the strong relativistic spherical shock in an ionized magnetized medium
is given by \citet{1977MNRAS.180..343B}. The  dynamics of both NR and ER shocks are governed
by the energy $E$ of the adiabatic impulsive (AI) explosion; while in the case of steady
injection of energy the dynamics is governed by $Lt$, where $L$ is the luminosity of
the central power supply. Other assumptions made by BM were that the shock is a strong
one (see Footnote \ref{strong}) and that the magnetic field is not dynamically important. 

The external medium into which the shock wave expands can have radial
variations of density. It can either be of uniform density
or could be stratified into $\rho\propto r^{-k}$, with $k=2$ for a constant velocity
wind. The shock can be either adiabatic or radiative. In adiabatic shocks
the radiative mechanisms are slow compared to hydrodynamics timescale, while
in radiative shocks the radiative mechanisms are faster than the hydrodynamic
timescale (measured in the observer-frame). A fully radiative blast 
wave may radiate away all the thermal energy
generated by the shock, if for example, the external medium is composed mainly
of electrons and positrons.
\citet{1998ApJ...509..717C} categorize shocks as ``semi-radiative" in which the
cooling is fast, but only a fraction of the energy is radiated away.  This could
take place in a collisionless shock acceleration, where the shock distributes the
internal energy between the electrons and protons. After the material passes 
behind the shock, there is no coupling, effectively at the low densities behind
the shock, between the electrons and protons. During the initial stage of the
afterglow, the electrons may undergo synchrotron cooling or Compton scatter off
low energy photons on timescales which are shorter than the dynamical timescale
\citep{1997ApJ...485L...5W,1998ApJ...499..301M,1998ApJ...497L..17S},
but the protons may not radiate and may remain hot. A narrow cooling layer may
form behind the shock as in a fast cooling scenario even when the protons remain
adiabatic and only the electrons cool. As the cooling parameter $\epsilon$ 
(the fraction of the energy flux lost in the radiative layer) increases, the
matter concentrates in a small shell near the shock in the Newtonian solution;
however in the ER case, even in the adiabatic case, the
matter is concentrated in a narrow shell of width $R/\Gamma^2$
(where $\Gamma=\sqrt{2}\gamma$, is the ultra-relativistic shock wave Lorentz factor
in the observer's frame), but this concentration in a narrower denser shell
increases with $\epsilon$.

Classical fireball models of GRBs have a fully radiative stage during the (initial)
$\gamma$-ray event with a radiative efficiency near unity and the implied energy
is typically $E \sim 10^{51-52} \; erg$ and
have a bulk Lorentz factor of $\Gamma \sim 10^2 - 10^3$. When the newly shocked electrons  
are radiating near the peak of the initial post-shock energy 
$\gamma_e \sim \xi_e (m_p/m_e) \Gamma(t)$ (here $\xi_e = O(1)$), they retain high radiative
efficiency even after the
GRB outburst for some time \citep{1998ApJ...499..301M}. If the protons establish and
remain in equipartition
with the electrons throughout the entire remnant volume, the shock Lorentz factor
evolves for a homogeneous external medium as: $\Gamma \propto r^{-3}$ 
\citep{1976PhFl...19.1130B}. On the other hand, for the adiabatic regime where the radiative
losses do not tap the dominant energy reservoir in protons and magnetic energy but only
the electrons are responsible for the radiation, one has: $\Gamma \propto r^{-3/2}$ for
a homogeneous medium \citep{1998ApJ...499..301M}.
 
For a NR blast wave of energy E into a medium of density $\rho_0$ 
in the Sedov-Taylor phase, it is
possible to construct a characteristic velocity
at a time t from the combination $(E/\rho_0 t^3)^{1/5}$. For the relativistic problem,
as BM point out,
an additional velocity the speed of light is introduced into the problem.
The mean energy per particle in the shocked fluid varies as $\Gamma^2$, where
one factor of $\Gamma$ arise from the increase in energy measured in the co-moving
frame and the second arise from a Lorentz transformation into the fixed (i.e. the observer)
frame. When most of the energy is resides in recently shocked particles, the total
energy is proportional to $\Gamma^2 R^3$.
Here R is the current shock radius. 
%
%
%
If the total energy contained in the shocked
fluid is to remain constant, then $\Gamma^2 \propto t^{-3}$.
One can consider a more general case in which the energy is supplied continuously at a rate
proportional to a power of the time and set:
\begin{equation}
\Gamma^2 \propto t^{-m} \text{ for } m>-1.
\end{equation}
%
For the m=3 case, corresponding to an impulsive injection of
energy into the blast wave on a time scale short compared with R, 
the total energy can be shown to be given by:
\begin{equation}
E = 8 \pi w_1 t^3 \Gamma^2 /17 ,
\end{equation}
where $w_1$ is the enthalpy ahead of the shock,
so that with $\Gamma^2 \propto t^{-3}$, the total energy is indeed a constant in time in the
fixed frame.

BM also provide a solution for an ER adiabatic
blast wave in an external density gradient $\rho_1 \propto  n_1  \propto r^{-k}$
but with a cold, pressure-less external medium. This
is the case for a blast wave propagating through a spherically symmetric wind. 
For this example they find that for k=2, m=1 corresponds to an impulsive energy
injection due to a blast wave in a constant velocity wind.
In this case, the analytic solution to the adiabatic impulsive blast waves can be
obtained for $m = 3 -k > -1$, and the total energy in this case is:
\begin{equation}
E = 8 \pi \rho_1 \Gamma^2 t^3 / (17-4k) .
\end{equation}
Again this energy is constant to the lowest order in $\Gamma^{-2}$.

We compare our solution to other well known analytic and numerical
schemes in the two final Figures. The graphs are not really for SN 2009bb-like
systems. As they are for a baryon loaded explosion with a $\gamma_0$ of 300 going
into a wind-like CSM. There may not be any such spherical explosions with these kind of
energies. The comparison in these figures are not for any
physically relevant objects (or scenarios) but rather for the purpose
of showing the consistency of our analytical solutions with previously
reported numerical techniques and known real asymptotic or intermediate
asymptotic solutions.

In Figure \ref{gamma_r} we compare our solution, in the ultra-relativistic
regime with nearly free expansion or the coasting solution at early times and a
BM like solution at intermediate times.
The blast wave solution for the special case of adiabatic impulsive solution for a constant
velocity wind obtained in Section \ref{blast} is valid
for arbitrary values of $M_0$ and $\gamma_0$ and the mass loss parameter $A$.
The corresponding
solutions in the Blandford McKee analysis can be obtained from
our exact solution in the limit of small initial mass of the high velocity
ejecta $M_0 \ll  m(R) \gamma_0$ where $m(R)$ is the swept-up mass in the wind. 
Thus Equation (B1) shows that our intermediate asymptotic expressions for $R$
and $\gamma$
have the same time dependence as those of the BM solution. Similarly,
the opposite limit $M_0 \gg  m(R) \gamma_0$ gives a nearly free expansion of the blast wave. 
Thus the relativistic CEDEX which had relatively low mass (although the prototype
SN 2009bb was still
significantly baryon loaded compared to classical GRBs),
underwent nearly free expansion initially. This is the relativistic analogue of the
\citet{1982ApJ...259..302C,1985Ap&SS.112..225N} phase of the non-relativistic
supernovae. Eventually, an ultra-relativistic CEDEX would enter the Blandford-McKee
phase (the relativistic analogue of the Sedov-Taylor phase) when it has swept up enough
mass from the external medium surrounding the CEDEX.

In Figure \ref{beta_r} we compare our solution, in the mildly-relativistic
regime with a BM like solution at intermediate times and a Snowplough phase at late times.
The BM-like phase becomes unphysical when $\gamma$ approaches 1 (hence $\beta$ becomes 0)
and the Snowplough phase is feasible only much later when $\beta$ is less than 1.
Both Figures also compare our solution to a direct numerical solution of the equations
of motion computed for this work, under under the relevant conditions as prescribed by \citet{1999ApJ...512..699C}.
These Figures demonstrate that patched solutions are much worse
than the solution proposed here or even numerical solutions. While there can be a patch
(though worse than our solution) between coasting and BM, there exists no patch between BM
and Snowplough. While our analysis is for a wind like ($\rho\propto r^{-2}$) medium,
see \citet{1999ApJ...513..669K} for similar transitions seen in numerical
simulations of the evolution of an adiabatic relativistic blastwave in an uniform medium.

Our framework may be used in
the future to study the transition of a relativistic blastwave
into a non relativistic Sedov phase.
\citet{2000ApJ...537..191F} have used the Sedov phase for very late time calorimetry
of GRB 970508.
The relation between the blast wave radius and the observer's frame time 
also leads to the appropriate numerical factor connecting the two
for a $\rho\propto r^{-2}$ exterior (See Appendix \ref{ultra}). Note
the different results in the literature due to
\citet{1997ApJ...489L..37S,1997ApJ...491L..19W} for an uniform media.


\section{Discussions}
\label{polao}
Explosion dynamics where adiabatic, ultra-relativistic or trans-relativistic
outflow takes place in a wind-like CSM can be classified in a parameter space
spanned by $M_0$, $\gamma_0$ and $A$,
i.e. the initial ejected mass, bulk Lorentz factor of the ejecta 
and the mass loss parameter of the external wind established before the explosion,
apart from the electron acceleration and magnetic field amplification efficiencies
parametrized by $\epsilon_e$ and $\epsilon_B$.
At low $\gamma_0$, we have the non relativistic supernovae, which usually have
considerable ejecta mass (e.g. at least $0.1 \; M_{\odot}$, often much larger). If the
progenitor of the SN has had very heavy mass loss soon before the explosion took place, then
an initially free expansion of the shock quickly sweeps up a mass equal to or greater than 
$M_0$ and the explosion enters the Sedov-Taylor like phase. Alternately a low ejecta mass 
initially would also lead to the early onset of the S-T phase.

The long duration GRB afterglows are powered by explosions that
have very small initial ejecta mass, typically $10^{-6} \; M_{\odot}$. Comparatively,
the initial mass of the ejecta in SN 2009bb estimated in Section \ref{inversion}
is considerably larger at $M_0 > 10^{-2.5} \; M_{\odot}$, but not as large as
non-relativistic supernovae. These explosions are therefore baryon loaded. As baryon loading
is a factor that affects the conversion of the impulsive release of energy into kinetic
energy of the matter around the central source, this can quench the emergence
of the gamma-rays in a burst. The substantial baryon loading in SN 2009bb compared
to classical GRBs may in fact have played a significant role in the circumstance that
no gamma-rays were in fact seen from this CEDEX, despite a thorough search by a
suite of satellites in the relevant time window. At the same time, the
CEDEX, such as SN 2009bb, are also unique in that
they span a region of $\gamma_0$ which is trans-relativistic. 
So far the LGRBs and the corresponding outflows
were being described in the extreme relativistic limit as in BM. Here we provide
a solution of the relativistic hydrodynamics equations which is uniquely tuned to the
CEDEX class of objects like SN 2009bb. Because of the non-negligible initial ejecta
mass of such a CEDEX, objects like this would persist in the free expansion phase for
quite a long time into their afterglow. Sweeping up a mass equal to that of the
original ejecta would take considerable time, unless the mass loss scale in its
progenitor was very intense, i.e. it had a large $A$.

In this paper we also provide an analysis of the peak radio flux versus the product
of the peak radio frequency and the time to rise to the radio peak. There are loci in this
plane that are spanned by low, intermediate and high velocity explosions.
Radio SNe, LGRBs and CEDEXs are seen to occupy different niches in this plane. Similarly,
we provide expressions for the peak fluxes and peak radio frequencies in terms of mass-loss
factors of the explosions. We also invert their dependence on $\gamma_0$ and $\dot{M}$
in terms
of the peak frequency, peak time and peak fluxes to interpret the parameters of the
explosion. We have also demonstrated that a seed magnetic field amplified
by the shock as described above plays a crucial role in the generation of synchrotron
radiation and have argued (Chakraborti et. al, in prep) that the acceleration
of cosmic rays to ultra-high energies is possible in SN 2009bb-like objects.



\acknowledgments

We thank Alicia Soderberg, Abraham Loeb and Poonam Chandra for discussions on
mildly relativistic supernovae. We thank Naveen Yadav for
checking the manuscript and Swastik Bhattacharya for help with mathematics. 
This work has made use of the Wolfram Mathematica computer algebra system.
This research was supported by the TIFR 11th Five Year Plan Project no. 11P-409.
We thank the Institute for Theory and Computation, Harvard University for its
hospitality. We thank an anonymous referee for detailed comments which
significantly extended this work, in particular for pointing out to us
the Chiang and Dermer papers, which we were unaware of. We thank
Charles Dermer for sharing their numerical results in these papers. We thank
Tsvi Piran and Charles Dermer for discussions at the Annapolis GRB2010 meeting.

\begin{figure}
\includegraphics[width=0.9\columnwidth]{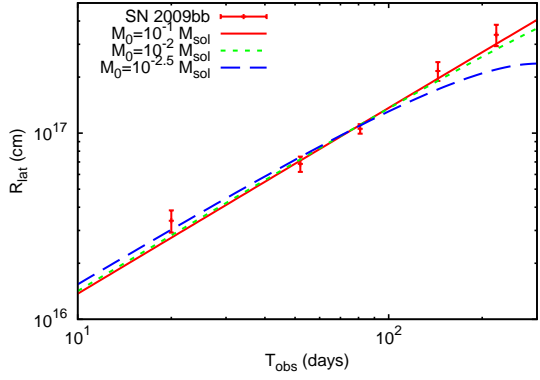}
\caption{Evolution of the blast wave radius $R_{lat}$, determined from SSA fit
to observed radio spectrum, as a function of the time $t_{obs}$ in the observer's frame.
The evolution is consistent with nearly free expansion. Note that the observations
require $M_0\gtrsim10^{-2.5}M_\odot$.\label{R_t}}
\end{figure}

\begin{figure}
\includegraphics[width=0.9\columnwidth]{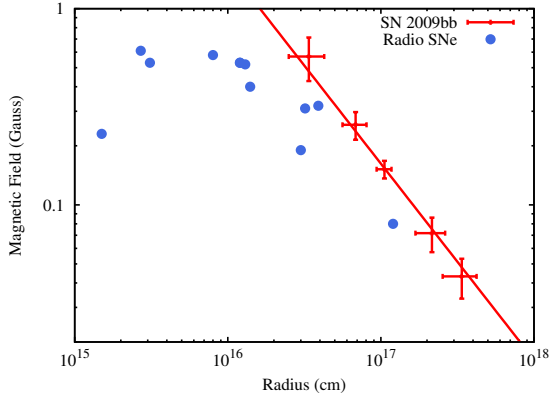}
\caption{Magnetic field as a function of blast wave radius, as determined from
SSA fits. Blue dots represent size and magnetic field of radio supernovae from
\citet{1998ApJ...499..810C} at peak radio luminosity. Red crosses (with $3\sigma$
error-bars) give the size and magnetic field of SN 2009bb at different epochs, from
spectral SSA fits. Red line gives the best $B\propto R^{-1}$ (Equation \ref{B})
fit.\label{BR}}
\end{figure}

\begin{figure}
\includegraphics[width=0.9\columnwidth]{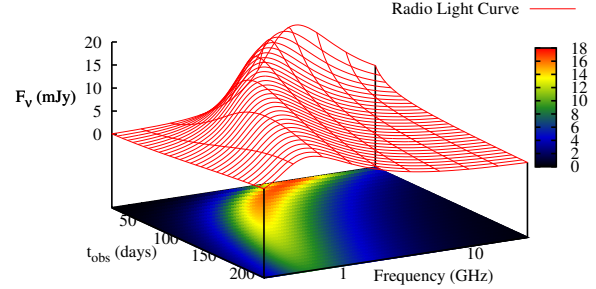}
\caption{The flux density $F_\nu$ as a function of the frequency $\nu$
and the time in the observer's frame $t_{obs}$, as predicted from the model
proposed in this work. The values used for the parameters, are those of
SN 2009bb.\label{lc}}
\end{figure}

\begin{figure}
\includegraphics[width=0.9\columnwidth]{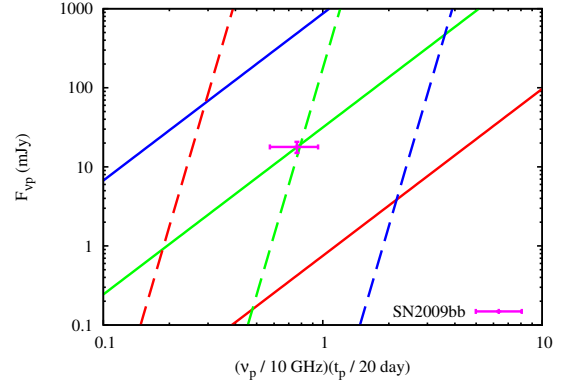}
\caption{Diagnostic $F_{\nu p}$ vs $\nu_p t_p$ plot of our fiducial CEDEX in
the nearly free expansion phase, at a distance
of 40 Mpc. Solid lines from right to left
represent loci of constant $\gamma_0=$ 1.005 (Red), 1.16
(Green) and 3.0 (Blue), traced by keeping the LHS of
Equation \ref{gamma0TV} fixed at the respective values.
Dashed lines from right to left
represent loci of constant $\dot{M}=$
0.2 (Red), 1.9 (Green) and 20 (Blue) in units of $10^{-6}{\rm~M_\odot yr^{-1}}$,
again traced by keeping the LHS of
Equation \ref{mdot0TV} fixed at the respective values.
The cross marks the set of values derived for the prototypical SN 2009bb
spectrum from \citet{2010Natur.463..513S} at 20 days, with 3$\sigma$ error-bars.
Radio observations can be used to deduce the initial bulk Lorentz factor $\gamma_0$
of a CEDEX and the pre-explosion mass loss rate $\dot{M}$ from this diagram.
\label{chev}}
\end{figure}

\begin{figure}
\includegraphics[width=0.9\columnwidth]{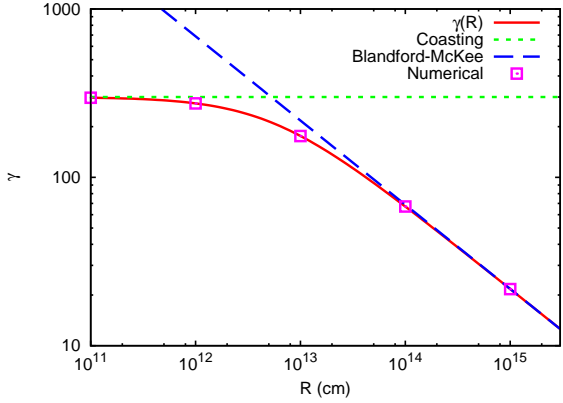}
\caption{
Early time radial evolution of the bulk Lorentz factor $\gamma(R)$ in a CEDEX,
derived in Equation \ref{gammaR} of this work (red solid line), as compared
to an initial nearly free expansion or coasting phase (green dotted line) and
the intermediate Blandford-McKee like phase (blue dashed line). Magenta squares represent
a direct numerical solution of the equations of motion following the method
prescribed by \citet{1999ApJ...512..699C}. Note the transition from a coasting phase
to the Blandford-McKee like phase.
\label{gamma_r}}
\end{figure}

\begin{figure}
\includegraphics[width=0.9\columnwidth]{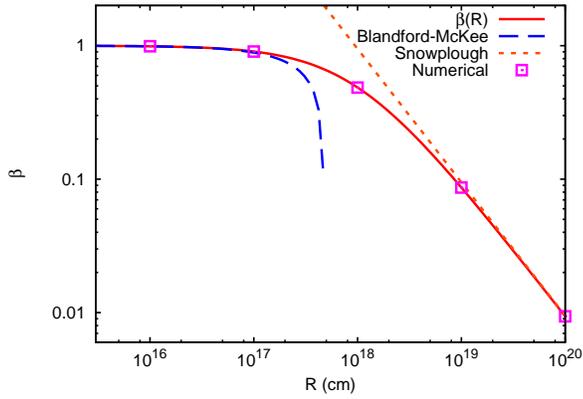}
\caption{
Late time radial evolution of $\beta(R)=v(R)/c$ in a CEDEX, derived using
Equation \ref{gammaR} of this work (red solid line), as compared to an
intermediate Blandford-McKee like phase (blue dashed line) and a final
Snowplough phase (orange dotted line). Magenta squares represent a direct
numerical solution of the equations of motion following the method prescribed
by \citet{1999ApJ...512..699C}. Note the transition from the Blandford-McKee
like phase to a Newtonian Snowplough phase.
\label{beta_r}}
\end{figure}

\clearpage

\appendix

\section{Temporal Evolution of Blastwave Parameters}
\label{exact}
The evolution of the blastwave radius has already been shown to be given by
\begin{equation}\label{Rt}
R=\frac{-M_0+2 A c \gamma_0 t+\sqrt{8 A c M_0 t \gamma_0^3+(M_0-2 A c \gamma_0 t)^2}}{2 A \gamma_0} .
\end{equation}
This can be substituted into into Equations (\ref{gammaR}) to get
\footnotesize
\begin{align}\label{Gt}
&\gamma=\left.\left(\gamma_0 M_0+\frac{-M_0+2 A c \gamma_0 t+\sqrt{8 A c M_0 t \gamma_0^3+(M_0-2 A c \gamma_0 t)^2}}{2 \gamma_0}\right)\right/ \\ \nonumber
&\sqrt{\left(2 A c \gamma_0 t+\sqrt{8 A c M_0 t \gamma_0^3+(M_0-2 A c \gamma_0 t)^2}\right)
   M_0+\frac{\left(-M_0+2 A c \gamma_0 t+\sqrt{8 A c M_0 t \gamma_0^3+(M_0-2 A c \gamma_0 t)^2}\right)^2}{4 \gamma_0^2}},
\end{align}
\normalsize
which gives the time evolution of the Lorentz factor $\gamma$. The amount of
kinetic energy converted into thermal energy $E$ is similarly obtained by
substituting $R(t)$ into Equations (\ref{ER}) to get
\footnotesize
\begin{align}
&E=c^2 \left(-M_0+\frac{M_0-2 A c \gamma_0 t-\sqrt{8 A c M_0 t \gamma_0^3+(M_0-2 A c \gamma_0 t)^2}}{2
   \gamma_0}\right. \\ \nonumber
&+\left.\sqrt{\left(2 A c \gamma_0 t+\sqrt{8 A c M_0 t \gamma_0^3+(M_0-2 A c \gamma_0 t)^2}\right)
   M_0+\frac{\left(-M_0+2 A c \gamma_0 t+\sqrt{8 A c M_0 t \gamma_0^3+(M_0-2 A c \gamma_0 t)^2}\right)^2}{4 \gamma_0^2}}\right)
\end{align}
\normalsize
Together, these three equations describe the temporal evolution of the blastwave
parameters. However, for times less than $t_{dec}$, it is sufficient to use the
first few Taylor coefficients, to derive the leading order behavior of the
observable quantities.

\section{Ultra-relativistic Limit for Negligible Ejecta Mass}
\label{ultra}
\citet{1976PhFl...19.1130B} describe the self similar evolution of an
ultra-relativistic blastwave in the $\gamma\gg1$ limit.
This is the relativistic analog of the non-relativistic
Sedov-Taylor
solution as the ejecta mass is considered negligible
compared to the swept up mass. \citet{1976PhFl...19.1130B} provide
a generalization of their solution to a blastwave plowing through a
$\rho\propto r^{-k}$ circumstellar profile. This is of particular interest for the
$k=2$ constant wind profile, relevant for the early evolution of some
GRB afterglows. For this situation, their solution provides a self similar
evolution with $R\propto t^{1/2}$ and $\gamma\propto t^{-1/4}$
\citep[see][Equation 77 with $k=2$]{2004RvMP...76.1143P}.

\citet{1972AnRFM...4..285B} have pointed out that self similar solutions
do not merely represent specific solutions, but describe the intermediate
asymptotic behavior of a wider class of solutions away from the boundaries
of the independent variables. Hence,
we investigate the low ejecta-mass limit ($\lim{M_0 \to 0}$) in our model,
keeping the initial energy $E_0=\gamma_0 M_0 c^2$ constant and obtain from
Equations (\ref{Rt} and \ref{Gt})
\begin{align}\label{BM}
&R\simeq \left(\frac{2E_0}{A c}\right)^{1/2} t^{1/2},
&\gamma\simeq \left(\frac{E_0}{2^3 A c^3}\right)^{1/4} t^{-1/4} .
\end{align}
Hence the ultra-relativistic Blandford-McKee solution for a $\rho\propto r^{-2}$
circumstellar media is a special limiting case of the general solution obtained
in this work. Our Equations (\ref{Rt} and \ref{Gt}) track the evolution of the blastwave
from nearly free expansion ($M_0\gg m(R) \gamma_0$) into the Blandford-McKee phase
($M_0\ll m(R) \gamma_0$). Note that unlike the self similar solutions
(eg. Equation \ref{BM}) there is no
divergence in quantities like the bulk Lorentz Factor at the $t=0$ boundary
of our solution (eg. Equation \ref{Gt}). As the result $\gamma_0$ is bounded above, by total energy
considerations.

Note that Equations (\ref{BM}) together
imply an unique relation between the blastwave radius and the time in the
observer's frame, $t_{obs}=R/(4 \gamma^2 c)$, which is different from the 
commonly used expression (Equation \ref{rees}) given by \citet{1997ApJ...476..232M}
by a factor of 2. The corresponding relation in case of a uniform external medium
has been argued over by \citet{1997ApJ...489L..37S,1997ApJ...491L..19W}.

\section{Stir-Fry Expressions: Radio Spectrum in Nearly Free Expansion Phase}
\label{radiotv}
The relativistic outflow in the prototypical SN 2009bb was discovered from
its strong radio emission \citep{2010Natur.463..513S}. Hence, more such
objects may be uncovered in radio follow up of type Ibc supernovae. 
Equations (\ref{nuptp} and \ref{Fnup}) completely describe the early
($t\lesssim t_{dec}$) temporal evolution of the radio afterglow of a CEDEX.
However, instead of expressing $F_{\nu p}$ and $\nu_p$ in CGS units in terms
of many constants, it would be of use to express them in units commonly used by
radio observers. Hence, substituting for the fundamental constants and replacing
for the appropriate choice of $c_1$, $c_5$ and $c_6$ from \citet{1970ranp.book.....P},
we express the peak flux density as
\begin{align}\nonumber
F_{\nu p} & \simeq 87\times\left(\frac{\epsilon_B}{0.33}\right)^{9/14}
\left(\frac{\epsilon_e}{0.33}\right)^{5/7} \left(\frac{f}{0.5}\right)^{-9/14}
(\gamma_0-1)^{19/14} \\
&\times\left(\frac{D}{40 {\rm~Mpc}}\right)^{-2}  \left(\left(\frac{\dot{M}}{10^{-6}
{\rm~M_\odot}}\right) \left(\frac{v_{wind}}{10^3 {\rm~km s}^{-1}}\right)^{-1}
\right)^{19/14}  {\rm~mJy}.
\label{FnupTV}
\end{align}
Here, the fiducial values of the parameters are chosen from those appropriate
for the prototypical SN 2009bb. Similarly, the temporal evolution of
the SSA peak frequency is given by,
\begin{align}\nonumber
\nu_p & \simeq 9.5\times\left(\frac{t_{obs}}{20 {\rm~days}}\right)^{-1}
\left(\frac{\epsilon_B}{0.33}\right)^{5/14}
\left(\frac{\epsilon_e}{0.33}\right)^{2/7} \left(\frac{f}{0.5}\right)^{-5/14} 
(\gamma_0^2-1)^{1/7} \\
& \times (\gamma_0+1)^{-9/14}
\left(\left(\frac{\dot{M}}{10^{-6}
{\rm~M_\odot}}\right) \left(\frac{v_{wind}}{10^3 {\rm~km s}^{-1}}\right)^{-1}\right)^{9/14} 
{\rm~GHz}.
\label{nupTV}
\end{align}
Given the model parameters, these equations together describe the peak
radio flux density of a CEDEX and the time evolution of its SSA peak in the
nearly free expansion phase. The flux density at any frequency $\nu$ and 
time $t$, can then be expressed as
\begin{equation}
F_\nu(t)\simeq 
\begin{cases} F_{\nu p}\left(\frac{\nu}{\nu_p(t)}\right)^{5/2} & \text{if $\nu<\nu_p$, or}\\
F_{\nu p}\left(\frac{\nu}{\nu_p(t)}\right)^{-(p-1)/2} &\text{if $\nu\geq\nu_p$,}
\end{cases}
\end{equation}
in terms of the already computed $F_{\nu p}$ and $\nu_p$, where an electron index
of $p\approx3$, is appropriate for a SN 2009bb-like spectrum with a optically thin
spectral index of $\alpha\approx-1$. These expressions also be should be useful in
designing radio surveys aimed at detecting CEDEXs.

\section{Extracting Blast Wave Parameters from Radio Observations}
\label{inversiontv}
The inverse problem is that of determining the initial bulk Lorentz factor
specified by $\gamma_0$, progenitor mass loss rate given by $A$ or $\dot{M}$
and the initial ejecta mass $M_0$, from the radio observations. 
The bulk Lorentz factor may be determined from the radio observations using
Equation (\ref{G0}) to get the simplified expression for $\gamma_0$ as,
\footnotesize
\begin{equation}
\gamma_0^2 \simeq 1+0.225\times
\left(\frac{\epsilon_B}{\epsilon_e}\right)^{2/19}
\left(\frac{f}{0.5}\right)^{-2/19} 
\left(\frac{t_{obs}}{20 {\rm~days}}\right)^{-2}
\left(\frac{\nu_p}{10 {\rm~GHz}}\right)^{-2}
\left(\frac{F_{\nu p}}{20 {\rm~mJy}}\right)^{18/19}
\left(\frac{D}{40 {\rm~Mpc}}\right)^{36/19}.
\label{gamma0TV}
\end{equation}
\normalsize
The result is insensitive to the equipartition parameter
$\alpha\equiv\epsilon_e/\epsilon_B$ and filling fraction $f$. This may be
used to reliably determine the initial bulk Lorentz factor of a radio detected
CEDEX in the mildly relativistic, nearly free expansion phase (like SN
2009bb).

An expression for $A\equiv\dot{M}/v_{wind}$, the circumstellar density profile,
set up by the mass loss from the progenitor may be obtained by eliminating $\gamma_0$
between equations (\ref{nuptp} and \ref{Fnup}). This gives us a complicated algebraic
dependence on $F_{\nu p}$ and $\nu_p t_p$. Since $\nu_p t_p$ is a pure number $\gg1$,
we can expand this expression in an asymptotic series \citep{Erdelyi} around
$\lim{\nu_p t_p \to \infty}$. This gives us the approximate expression for the mass
loss rate as
\begin{align}\nonumber
\dot{M} & \simeq 3.0\times10^{-6}
\left(\frac{\epsilon_B}{0.33}\right)^{-11/19}
\left(\frac{\epsilon_e}{0.33}\right)^{-8/19}
\left(\frac{f}{0.5}\right)^{11/19}
\left(\frac{v_{wind}}{10^3 {\rm~km s}^{-1}}\right)^{1} \\
&\times \left(\frac{t_{obs}}{20 {\rm~days}}\right)^{2}
\left(\frac{\nu_p}{10 {\rm~GHz}}\right)^{2}
\left(\frac{F_{\nu p}}{20 {\rm~mJy}}\right)^{-4/19}
\left(\frac{D}{40 {\rm~Mpc}}\right)^{-8/19}
{\rm~M_\odot yr^{-1}}.
\label{mdot0TV}
\end{align}
This approximate expression indicates the dependence of the inferred mass loss rate
on the observational parameters, and makes an error of only $\lesssim10\%$ in case of
SN 2009bb, when compared to the exact expression. Given the uncertainties in the
observations, we recommend the use of this expression to get an estimate of the
mass loss rate from a CEDEX progenitor. Note that, this expression has similar
scaling relations as Equation (23) of \citet{2006ApJ...651..381C}. Hence, the
mass loss rate of SN 2009bb as determined using that equation by
\citet{2010Natur.463..513S} remains approximately correct.

The initial ejecta rest mass $M_0$ cannot be estimated from radio observations
in the nearly free expansion phase. It can
only be determined when the CEDEX ejecta slows down sufficiently due to interaction
with the circumstellar matter (Figure \ref{R_t}). Thereafter, the initial ejecta
mass can be obtained from the Equation (\ref{dec}) using the timescale of slowdown
$t_{dec}$ and the already determined $A$ and $\gamma_0$. Nearly free expansion for a
particular period of time, can only put lower limits on the ejecta mass, as shown
in this work.

\section{Angular Size Evolution and VLBI}
\label{vlbi}
Very Long Base Interferometric (VLBI) measurements of the apparent angular
diameters may be compared to the predicted value of $R_{lat}$, as a
direct test of the CEDEX model. In the nearby universe, where the luminosity
distance and the angular distance are not significantly different, we can
substitute Equation (\ref{gamma0TV}) into Equation (\ref{RLS}) to obtain
the predicted angular diameter $\theta\simeq2R_{lat}/D$ of a CEDEX as,
\begin{equation}
\theta \simeq 82\times
\left(\frac{\epsilon_B}{\epsilon_e}\right)^{1/19}
\left(\frac{f}{0.5}\right)^{-1/19} 
\left(\frac{\nu_p}{10 {\rm~GHz}}\right)^{-1}
\left(\frac{F_{\nu p}}{20 {\rm~mJy}}\right)^{9/19}
\left(\frac{D}{40 {\rm~Mpc}}\right)^{-1/19}
{\rm~\mu as}.
\label{thetaTV}
\end{equation}
This expression which is insensitive to the equipartition parameter, the filling
factor and even the assumed distance to the source, may be used in planning
VLBI observations of a radio detected CEDEX and for comparing the observed angular
sizes with those predicted from our model.

At $t_{obs}\simeq81$ days post explosion, the radio spectrum of SN 2009bb, constructed
from broadband Giant Metrewave Radio Telescope (GMRT) and Very Large Array (VLA)
observations, as given by \citet{2010Natur.463..513S} is well fit by an SSA spectrum
with $F_{\nu p}=10.82\pm0.34$ mJy and $\nu_p=1.93\pm0.07$ GHz. Substituting these values
into Equation (\ref{thetaTV}) we have the angular diameter as
$\theta=318\pm12 \mu$as, for the fiducial values of $f$, $\epsilon_B$ and
$\epsilon_e$ adopted in this work. Hence, the angular radius of $0.16$ mas
predicted by our model is consistent with the $3\sigma$ upper limit
\citep{2010arXiv1006.2111B} of $0.64$ mas reported from VLBI observations at
around $t_{obs}\simeq85$ days. \citet{2010arXiv1006.2111B} adopt a Sedov-Taylor
expansion for the blastwave, according to which the source will not be resolved
anytime soon. However, our model indicates a nearly free, mildly relativistic
expansion for SN 2009bb, which may soon be resolvable at the VLBI scale. We note, by
$t_{obs}\simeq222$ days, the radio spectrum had evolved to $F_{\nu p}=8.35\pm0.59$ mJy
and $\nu_p=0.53\pm0.04$ GHz, predicting an angular radius of around $0.51$ mas
according to Equation (\ref{thetaTV}). We therefore predict that a
careful VLBI observation will now successfully resolve the radio
emission from SN 2009bb and confirm the nearly free expansion indicated
by our analytic solution. Given the lower fluxes at the usually high VLBI
frequencies, this may be a challenging observation due to inadequate sensitivity and
unsuitability of self-calibration techniques at low fluxes.





\bibliographystyle{apj}
\bibliography{rel_sne}




\end{document}